\documentclass[twocolumn,showpacs,showkeys,preprintnumbers,amsmath,amssymb,aps]{revtex4}

\usepackage{graphicx}
\usepackage{color}
\usepackage{float}

\begin{document}

\preprint{Barth et al., Co$_2$TiZ}

\title{Anomalous transport properties of the halfmetallic ferromagnets Co$_2$TiSi, Co$_2$TiGe, and Co$_2$TiSn}

\author{Joachim Barth, Gerhard H. Fecher, Benjamin Balke, Tanja Graf,
        and Claudia Felser}
\email{felser@uni-mainz.de}
\affiliation{Institut f\"ur Anorganische und Analytische Chemie, \\
             Johannes Gutenberg - Universit\"at, D-55099 Mainz, Germany.}

\author{Andrey Shkabko and Anke Weidenkaff}

\affiliation{EMPA, Swiss Federal Laboratories for Materials Testing and Research,
             Solid State Chemistry and Catalysis, CH-8600 Duebendorf, Switzerland}

\date{\today}

\begin{abstract}

In this work the theoretical and experimental investigations
of Co$_{2}$Ti$Z$ ($Z$ = Si, Ge, or Sn) compounds are reported.
Half-metallic ferromagnetism is predicted for all three compounds
with only two bands crossing the Fermi energy in the majority channel. 
The magnetic moments fulfill the Slater-Pauling rule and the Curie temperatures
are well above room temperature. All compounds show a metallic like resistivity
for low temperatures up to their Curie temperature, above the resistivity
changes to semiconducting like behavior. 
A large negative magnetoresistance of 55~\% is observed for 
Co$_{2}$TiSn at room temperature in an applied magnetic field of $\mu_0H=4~T$
which is comparable to the large negative magnetoresistances of the manganites.
The Seebeck coefficients are negative for
all three compounds and reach their maximum values at their respective
Curie temperatures and stay almost constant up to 950~K. The highest value achieved is -52~$\mu$VK$^{-1}$m$^{-1}$ for Co$_{2}$TiSn which
is large for a metal. The combination of half-metallicity and the
constant large Seebeck coefficient over a wide temperature range makes these compounds
interesting materials for thermoelectric applications and further spincaloric investigations.

\end{abstract}



\maketitle


In the last few years Heusler alloys have attracted a lot of interest
as suitable materials for spintronic applications~\cite{FFB07}.
A huge amount of theoretical and experimental studies investigating the half-metallic properties
and enhancing the performance
of the compounds and devices for different types of applications were
done. In 2005, Hashemifar {\it et al}~\cite{HKS05} showed in a
density functional theory study the possibility to preserve
the half-metallicity at the surface of the Heusler compound Co$_2$MnSi(001). 
Very recently, Shan {\it et al}~\cite{SSW09} demonstrated experimentally
the half-metallicity of Co$_2$FeAl$_{0.5}$Si$_{0.5}$ at room temperature.
The ability of growing well ordered half-metallic Heusler thin films on
semiconductors (e.g. Ge (111)) makes Heusler compounds suitable
for spininjection as shown by Hamaya {\it et al}~\cite{HIN09}.
The transport properties of pure and doped Fe$_2$VAl show semiconductor like
behavior with interesting and anomalous features near the
structural transition temperature with an increasing absolute Seebeck
coefficient with increasing temperature~\cite{NKA97,LCL07}.
The recent observation of the spin Seebeck effect
allows to pass a pure spin current over a long distance \cite{UTH08}
and is directly applicable to the production of
spin-voltage generators which are crucial for driving spintronic devices~\cite{WAB01, ZFS04, CFV07}.
To generate effectively a spin current and for other spincaloric applications as well
one needs half-metals with a constant Seebeck effect behavior~\cite{Ong08}.
A constant Seebeck effect in dependence of the temperature is found
especially in correlated systems such as pure and doped SrRuO$_3$~\cite{KHM06}.
It can also be achieved e.g. in PbTe crystals with a sophisticated and experimentally
complex graded Indium doping through the crystal~\cite{DSD02}.
This report will show that the Heusler compounds Co$_2$TiSi, Co$_2$TiGe, and Co$_2$TiSn
combine all requirements for the demand of half-metallic ferromagnets
with a constant behavior of the Seebeck coefficient to be used
in future for spincaloric devices. A big advantage is the easy tunability 
of the properties of the Heusler compounds which makes it easy to design 
new materials ~\cite{BWF08, BFG08}.

\begin{figure}
\centering
\includegraphics[width=8cm]{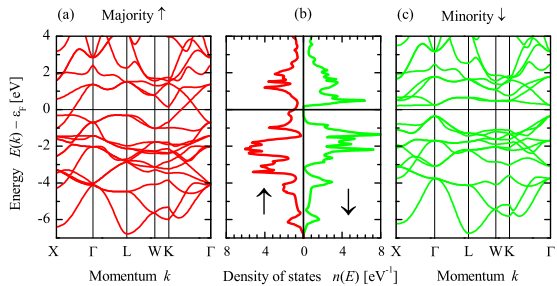}
\caption{Spin resolved electronic structure of Co$_2$TiGe. \\
         Shown are the band structures for the majority (a)
         and minority (c) electrons together with the density of states (b).}
\label{Fig1_BS-DOS_new}
\end{figure}

The electronic structure of the compounds was calculated by means of
the full potential linearized augmented plane wave (FLAPW)
method~\cite{BSS90,BSM01}. Details of the calculations are reported
in Ref.~\cite{KFF07}. The calculated band structure of Co$_2$TiGe is shown
in Figure~\ref{Fig1_BS-DOS_new}. A band gap in the minority states occurs
at the Fermi energy $\epsilon_F$. This minority gap is characteristic
for half-metallic materials as was previously shown by various
authors~\cite{MAH06,KFF07,IAK82,GDP02,LLB05}. The size of the minority
band gap amounts to $\Delta E_{\rm{HMF}}=500$~meV and the Fermi energy
is located close to the minimun of the conduction band. 
This was just confirmed by the experimental investigation
of the spin-resolved unoccupied density of states \cite{KKB09}.
The majority band structure is metallic with only two bands crossing 
$\epsilon_F$ and an electronic instability
(van-Hove singularity) close to the $L$ point just below $\epsilon_F$.
Due to the band gap in the minority states the magnetic
moment is integer and has a value of exactly 2~$\mu_B$ for the
primitive cell. A closer analysis shows that the magnetic moment is
located at the Co atoms only. That means, each of the two Co atoms in
the primitive cell carries a magnetic moment of 1~$\mu_B$.

The Co$_2$TiZ samples were produced by arc melting of stoichiometric amounts of the
elements and afterwards annealed in evacuated quartz tubes for 21~days.
For more details about the sample preparation see Ref.~\cite{BFK06}.
All samples exhibit the L2$_{1}$ structure and the lattice parameters at 300~K
are determined to $a=5.733$~{\AA}, $a=5.819$~{\AA}, and $a=6.066$~{\AA}
for Co$_2$TiSi, Co$_2$TiGe, and Co$_2$TiSn, respectively.
The obtained values are in good agreement with previously published values~\cite{EBE83, WeZ83, BEJ83}.

The magnetic properties of the Co$_2$TiZ samples were investigated by a super
conducting quantum interference device (SQUID, Quantum Design
MPMS-XL-5). The Co$_2$-based Heusler alloys that are
half-metallic ferromagnets show a Slater-Pauling like behavior for the
magnetization, what means that the saturation magnetization scales
linearly with the number of valence electrons~\cite{GDP02,FKW06}. This
results in a theoretical magnetic moment of 2~$\mu_B$ per formula unit
at T=$0$~K for all three compounds.

\begin{figure}
  \centering
  \includegraphics[width=8cm]{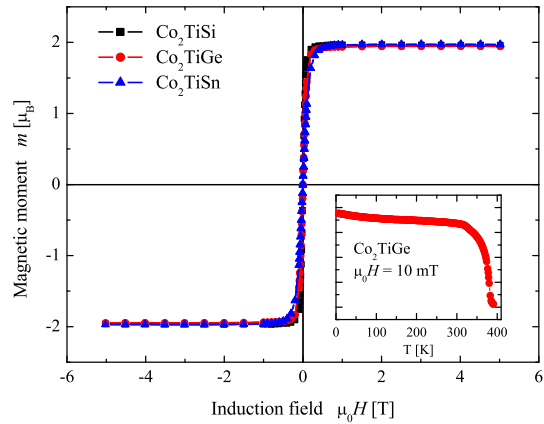}
  \caption{Magnetization measurements of Co$_2$TiZ.\newline
         Displayed are the hysteresis curves at 5K for
         Co$_2$TiSi, Co$_2$TiGe, and Co$_2$TiSn.
         The inset shows the temperature dependence measured in an
         induction field of $\mu_0H=10mT$ for Co$_2$TiGe.}
  \label{Fig2_SQUID_Co2TiZ}
\end{figure}

The saturation magnetization data measured at 5~K are shown in Figure~\ref{Fig2_SQUID_Co2TiZ}
and reveal that all three compounds
fulfil the Slater-Pauling rule and therefore are potential half-metallic ferromagnets.
The magnetic moments at 5~K are 1.96~$\mu_B$, 1.94~$\mu_B$, and 1.97~$\mu_B$ per formula unit
for Co$_2$TiSi, Co$_2$TiGe, and Co$_2$TiSn, respectively.
The Curie temperature $T_C$ was determined from the temperature dependence of the
magnetisation measured in an induction field of $\mu_0H=10mT$ (see inset of Figure~\ref{Fig2_SQUID_Co2TiZ}).
The values for $T_C$ are 380~K for Co$_2$TiSi and Co$_2$TiGe, and 355~K for Co$_2$TiSn .

The measurements of the transport and thermoelectric properties were carried out with
a Physical Property Measurement System (Model 6000 Quantum Design)
on bars of about $(2\times2\times8)$~mm$^3$ which were cut from the pellets and polished
before the measurement.
The resistivity data for temperatures from 2~K to 400~K were obtained by a
standard AC four probe method and the data from 350~K to 950~K were
measured by a standard DC four point method.

\begin{figure}
  \centering
  \includegraphics[width=8cm]{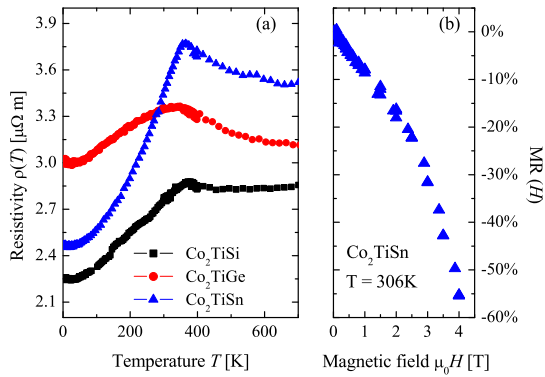}
	\caption{ Measured temperature dependent electrical resistivity $\rho$ of Co$_{2}$TiZ
	          compounds with Z = Si, Ge, or Sn (a). Magnetoresistance as a function 
	          of the applied magnetic field for Co$_{2}$TiSn at $T$=306~K (b).}
\label{Fig3_Resistivity_Co2TiZ}
\end{figure}

Figure~\ref{Fig3_Resistivity_Co2TiZ}(a) shows the temperature dependence of the
resistivity $\rho$ for Co$_2$TiSi, Co$_2$TiGe, and Co$_2$TiSn.
The resistivity behavior is metallic for all compounds in the
low temperature range from 2~K to the respective Curie temperature. At
$T_C$ a cusp-type transition is observed~\cite{Kat01}. From there
the resistivity declines to 550~K and stays nearly constant at higher
temperatures. The Curie temperatures can be estimated from the maxima of the
resistivities. The determined values are 370~K, 350~K, and 360~K for
Co$_2$TiSi, Co$_2$TiGe, and Co$_2$TiSn, respectively. The values agree
well with the ones obtained from the magnetic measurements.
The measurements agree well with previous
findings of other investigations for temperatures below 450~K~\cite{MCS05}.
Above $T_C$, a decrease of the resistivity is observed. The decrease is
very pronounced up to about 600~K and stays nearly constant at higher
temperatures. This behavior leads to a significant anomaly of the
resistivity at $T_C$. Other than in many ferromagnetic materials, here
the Curie temperature can be easily related to the maximum of the
resistivity.

Similar, cusp-type anomalies of the resistivity close to $T_C$ 
are very typical for the manganites which show a large negative 
magnetoresistance (MR) around $T_C$~\cite{HWH93, SRB95}. This was observed in 
GdI$_2$ as well which shows a large room temperature MR of around 70~\% at 7~T~\cite{FAK99}.
For intermetallic compounds these anomalies were only observed for Heusler
compounds~\cite{HAC81, NIA93}.
In Figure~\ref{Fig3_Resistivity_Co2TiZ}(b) the magnetoresistance of
Co$_{2}$TiSn at $T$=306~K is shown as defined by $[R(B)-R(0~Oe)]/R(0~Oe)$.
In an applied magnetic field of $\mu_0H=4~T$ the MR exceeds 55~\% which is an enormous
value for a polycrystalline sample at room temperature.

\begin{figure}
  \centering
  \includegraphics[width=8cm]{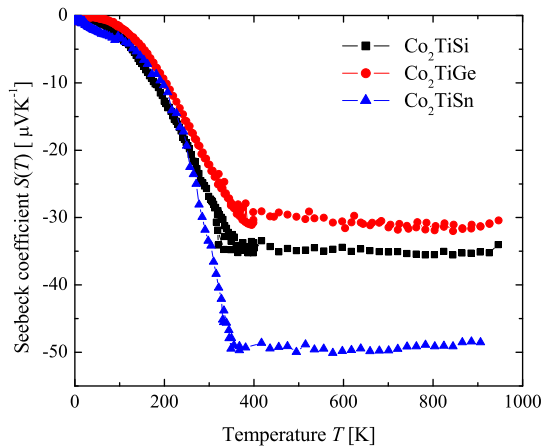}
	\caption{Measured Seebeck coefficient of Co$_{2}$TiZ compounds
	         with Z = Si,  Ge, or Sn.}
	\label{Fig4_Seebeck_Co2TiZ}
\end{figure}

Figure~\ref{Fig4_Seebeck_Co2TiZ} shows the Seebeck coefficient (S) that was
measured for temperatures from 2~K to 950~K. The values decrease with
increasing temperature up to the Curie temperature and stay almost
constant at higher temperature. The mean values above $T_C$ are
-35~$\mu$VK$^{-1}$, -31~$\mu$VK$^{-1}$, or -51~$\mu$VK$^{-1}$ for
Co$_2$TiSi, Co$_2$TiGe, and Co$_2$TiSn, respectively. The maximum
absolute values are rather large compared to the most elemental metals
and are close to the Seebeck coefficient of elemental Co that exhibits
about -30~$\mu$VK$^{-1}$ to -45~$\mu$VK$^{-1}$ at temperatures between
300~K and 500~K.

The Seebeck coefficient of all three investigated compounds is negative
over the entire temperature range. For Co$_2$TiSn a negative increase 
of the Seebeck coefficient is observed at about 80~K. This is close to 
$T / \Theta_D \approx 0.2$ with $\Theta_D$ being the Debye temperature
where the largest influence of a phonon drag is expected.
This assumption is in agreement with the measurement of the thermal 
conductivity where a peak at low temperatures is observed.
The Seebeck coefficient provides a sensitive test of the electronic
structure for metals in the vicinity of the Fermi energy.
The rather large constant Seebeck coefficients above the Curie temperatures up to at least 950~K
could be achieved by filling up the flat band in the $\Delta$ direction
in the minority spin channel just above $\epsilon_F$ (see Figure~\ref{Fig1_BS-DOS_new}(c)).
The constant Seebeck coefficients above the Curie temperatures
mean that the energy of the Fermi level is pinned in a wide temperature range in these compounds~\cite{DSD02,LYW05}
and therefore the thermovoltages have a linear temperature dependence which
makes these compounds suitable materials for future thermoelectric devices
such as thermocouples. With an Al-doping of the $Z$-Position in the Co$_2$Ti$Z$ ($Z$ = Si, Ge, or Sn) compounds
one can tune their Curie temperature to lower values~\cite{GFB09} and therefore tune the working temperature
of thermocouples designed out of these materials.
In cubic systems the Seebeck coefficient is derived from the scalar 
electrical resistivity $\rho$ and thermal conductivity $\nu$: 
$S =\rho \times \nu$. Therefore, the constant value of the Seebeck effect means that the product of $\rho$
and $\nu$ have to be constant so they have to compensate each others dependency.




In summary, the Co$_{2}$TiZ (Z = Si, Ge, or Sn) compounds were investigated
theoretically and experimentally. Band structure calculations predict
half-metallic ferromagnetism for all three compounds.
The size of the minority band gap amounts to $\Delta E_{\rm{HMF}}=500$~meV and the Fermi energy
is located close to the middle of the gap. The majority band structure
is metallic with a single band crossing $\epsilon_F$ in the $\Delta$
direction.
The Curie temperature and the magnetic moment
were determined by SQUID measurements and reveal that all three
compounds fulfill the requirement for half-metallicity according
to the Slater-Pauling rule and have Curie temperatures
well above room temperature.
All compounds show a metallic like resistivity
for low temperatures up to their Curie temperature. From there on they
change to semiconducting like behavior. This behavior is attributed to
a ferromagnetic to paramagnetic transition, that strongly influences
the band structure. The Seebeck coefficients are all negative and reach
their maximum values at their respective Curie temperatures. The
highest value achieved is -52~$\mu$VK$^{-1}$m$^{-1}$ for Co$_{2}$TiSn.
Furthermore, the Seebeck coefficients of all
three compounds are rather large for metals and constant above the Curie temperatures. 
Therefore the temperature dependence of the thermovoltages is linear
which ensures a stable conversion efficiency even at high temperatures.
This makes the compounds very attractive as materials for thermocouples or thermoelectric generators.
The combination of half-metallicity and the
constant large Seebeck coefficient over a wide temperature range makes these compounds
interesting materials for further spincaloric and thermoelectric investigations.
This work is financially supported by "Stiftung Innovation
Rheinland-Pfalz", the Deutsche Forschungsgemeinschaft DFG (projects
P~01 and P~07 in research unit FG~559) and DAAD (D06/33952).

\bigskip

\bibliography{Co2TiZ_PRB}

\end{document}